\documentclass[3p,times,twocolumn]{elsarticle}

\newcommand\bx{\mathbf{x}}

\newcommand\atoprep[2]{\genfrac{}{}{0pt}{}{#1}{#2}}

\usepackage[colorlinks]{hyperref}

\usepackage{amssymb}
\usepackage{mathtools}
\usepackage{booktabs}
\usepackage{siunitx}

\usepackage{lineno}

\biboptions{sort&compress}

\journal{arXiv}

\graphicspath{{./figures/}}

\begin{document}

\begin{frontmatter}



\title{Structural Characterization of Grain Boundaries and Machine Learning of Grain Boundary Energy and Mobility}


\author[byuP]{Conrad W. Rosenbrock}
\author[byuME]{Jonathan L. Priedeman}
\author[byuP]{Gus L. W. Hart}
\author[byuME]{Eric R. Homer}


\address[byuP]{Department of Physics and Astronomy, Brigham Young University, Provo, UT 84602, USA}
\address[byuME]{Department of Mechanical Engineering, Brigham Young University, Provo, UT 84602, USA}

\begin{abstract}
Recent advances in the numerical representation of materials opened the way for successful machine learning of grain boundary (GB) energies and the classification of GB mobility and shear coupling. Two representations were needed for these machine learning applications: 1) the ASR representation, based on averaged local environment descriptors; and 2) the LER descriptor, based on fractions of globally unique local environments within the entire GB system. We present a detailed tutorial on how to construct these two representations to learn energy, mobility and shear coupling. Additionally, we catalog some of the null results encountered along the way.





\end{abstract}


\end{frontmatter}



\section{Introduction}

Many material behaviors depend crucially on the structure and properties of grain boundaries (GBs). Unfortunately, predicting the properties of the vast array of possible GBs that result from their atomic structures represents a particular challenge. This work goes into detail on recent efforts \cite{Rosenbrock2017,Priedeman:2018structure} to characterize the atomic structure of GBs and infer GB structure-property relationships using machine learning.

The key component to any machine learning solution is the representation of the data to be learned \cite{bartok2013representing,szlachta2014accuracy,de2015comparing}. Specifically, in the case of GBs, a good descriptor would have the following properties: 1) be agnostic to the grains’ specific underlying lattices (or lack of periodicity); 2) have invariance to global translation, global rotation, and permutations of identical atoms; 3) lead to a metric that is smooth and differentiable; 4) provide a quantifiable comparison of (dis)similarity between different structures. Ideally, a small change in the atomic positions of some atoms near a GB should register a correspondingly small change in the descriptor and metric between the original and perturbed boundaries. The ability of the descriptor/metric combination to distinguish the (dis)similarity of the GBs limits the machine’s ability to learn the underlying science.

In this work, we present in detail a new descriptor for single-species GBs based on the Smooth Overlap of Atomic Positions (SOAP) descriptor \cite{PhysRevLett.104.136403,bartok2013representing}. This descriptor represents each local atomic environment (LAE) within a GB as a numerical vector. Once the matrix of descriptors is constructed for each GB in the system, we can combine them to form a single matrix for the entire GB system either 1) by averaging to form the Averaged SOAP Representation (ASR); or 2) by finding the subset of LAEs in the GB system that are unique, and then representing a GB by its fractions of each unique LAE to form the Local Environment Representation (LER).

In the method details section, we describe how to construct a SOAP matrix for a given GB and how to form the ASR and LER for the GB system as well as the influence of different parameters on the LER. We also explain how to predict the grain boundary energy, mobility and shear coupling for a set of Nickel GBs using these two representations and machine learning algorithms. The additional details section provides a few comments and concluding remarks.


\section{SOAP descriptor for local atomic environments}
\label{sec:SOAP_ASR_LER}

To develop our descriptors of any LAE, we apply the SOAP formalism  \cite{PhysRevLett.104.136403,bartok2013representing}.
We define the species-independent neighbour density of atom $i$ as
\begin{eqnarray}
\label{eq:NeighborDensityDef}
  \rho_{is}(\vec{r}) & = & \sum_{j} e^{-(\vec{r}_{ij}-\vec{r})^2/2\sigma_\mathrm{atom}^2}
                           f_\mathrm{cut}(|\vec{r}_{ij}|)
\end{eqnarray}
where $f_\mathrm{cut}$ is a smooth cutoff function that ensures compact
support, and $\vec{r}_{ij}$ is the vector from atom $\vec{r}_i$ to
$\vec{r}_j$. This is equivalent to placing a Gaussian at each atomic
position in the local neighborhood of the atom.

The overlap of two different site environments is defined as
\begin{eqnarray}
\label{eq:OverlapS}
  S(\rho_i, \rho_k) & = & \int \rho_i(\vec{r}) \rho_k(\vec{r}) d^3r.
\end{eqnarray}

This overlap is permutationally invariant (because of the sum in
$\rho_{i}$), but not yet rotationally invariant. In order to make it
so, we integrate it over all rotations of one of its arguments.
\begin{equation}
\tilde{K}(\rho_i, \rho_k) = \int d\hat R \,|S(\rho_i, \hat R\rho_k)|^p,
\end{equation}
where $\hat R$ is a 3D rotation operator (element of SO(3)), and $p$
is a small integer, e.g., 2. Finally the normalised SOAP kernel
(descriptor) is
\begin{equation}
\label{eq:FullKernel}
  K(\rho_i, \rho_k) = \frac{\tilde K(\rho_i, \rho_k)}{\sqrt{\tilde
      K(\rho_i, \rho_i) \tilde K(\rho_k,\rho_k)}}.
\end{equation}

Now we derive the efficient formula to evaluate the SOAP kernel. We
start by expanding the neighbour density for each species in an
orthonormal basis,
\begin{equation}
\label{eq:RhoExpansionPSNL}
\rho_{i}(\vec{r}) = \sum_{nlm} c_{i,nlm} \, g_n(r) Y_{lm}(\hat r), 
\end{equation}
where $g_n$ are an orthonormal radial basis, $Y_{lm}$ are spherical
harmonics, and $c_{i,nlm}$ are the expansion coefficients. The effect
of the rotation operator acting on $\rho_{k}$ is written in terms of
Wigner matrices,
\begin{equation}
\hat R\rho_{k}(\vec{r}) = \sum_{nlmm'} D^l_{mm'}(\hat R) c_{k,nlm'} \,g_n(r)Y_{lm}(\hat r).
\end{equation}
\noindent So the overlap $S$ is given by
\begin{equation}
S(\rho_i, \hat{R} \rho_k) = \sum_{nlmm'} c_{i,nlm}^{*} D^l_{mm'}(\hat R) c_{k,nlm'}.
\label{overlap}
\end{equation}

We now need to square this expression and integrate over all
rotations. This is aided by the formula
\begin{equation}
\int d\hat R \,D^l_{mm'}(\hat R)D^\lambda_{\mu\mu'}(\hat R) =
\delta_{l\lambda}\delta_{m\mu}\delta_{m'\mu'} \frac{1}{\sqrt{2l+1}},
\end{equation}

which leads to the unnormalised kernel
\begin{equation}
\label{eq:UnnormalizedK}
\tilde K(\rho_i,\rho_k)= \sum_{nn'\atoprep lmm'}
\frac{c_{i,nlm}^{*}c_{k,n'lm'} c_{i,n'lm}^{*}
  c_{k,n'lm'}}{\sqrt{2l+1}}.
\end{equation}

We can write this kernel as a dot product (called the ``power
spectrum'') of rotationally invariant descriptors of the two
environments. In order to simplify the notation, we also introduce
the set of indices $\bx \equiv nn'l$ and $\bx' \equiv n''n'''l''$;
thus the expression $\sum_{\bx} \equiv \sum_{nn'l}$, etc. Using
\begin{equation}
\label{eq:pVectorDefinition}
p_{i,\bx} = \sum_{m=-l}^l c_{i,nlm}^{*} c_{i,n'lm},
\end{equation}
we can now write the full, un-normalized kernel as
\begin{equation}
\label{eq:FinalTildeK}
\tilde K(\rho_i, \rho_k) = \sum_{\bx \bx'} p_{i,\bx}\, p_{k,\bx'}.
\end{equation}

Inserting this definition into Eq. (\ref{eq:FullKernel}) produces a
value between 0 and 1 defining the similarity between two local atomic
environments. The tunable parameters are the local neighborhood cutoff
value (used by $f_{\mathrm{cut}}$), and the expansion limits for the
radial and angular basis functions $n_{\mathrm{max}}$ and
$l_{\mathrm{max}}$ in Eq. (\ref{eq:RhoExpansionPSNL}).

An efficient implementation of this procedure is available in the Python-based QUIP
code (http://www.libatoms.org/)
\cite{PhysRevLett.104.136403,bartok2013representing}. It includes an
interface similar to that of ASE \cite{ISI:000175131400009} so that
SOAP descriptors can easily be calculated for any collection of
atoms. Thus, the problem of generating a SOAP descriptor for a GB
reduces to the procedure for selecting those atoms in the crystal that
should be considered part of the GB.

\section{Grain boundary descriptor}

\label{sec:GBAtomSelection}

In applying the descriptor in Eq. (\ref{eq:FullKernel}) to grain
boundaries, we first need to isolate the atoms at the boundary. 
In our original work \cite{Rosenbrock2017}, we
knew \emph{a priori} that we wished to learn energies from the Averaged
SOAP Representation (ASR) and classify mobility and shear coupling
using the Local Environment Representation (LER). This informed our
selection of which atoms to include. Since ASR includes a sum over
\emph{all} the atoms' local environment descriptors, including a large
volume of well-ordered (i.e., perfect crystal) atoms in the grain, it
dilutes the important deviations contributed by LAEs at the
boundary. For the ASR then, we include only those atoms with greatest
structural deviation at the boundary and then introduce an artificial
surface to isolate these GB atoms from the rest of the bulk. Since
energy is an average property of the entire GB, this artificial
surface does not interfere with the interpretability of our results.

For the LER, on the other hand, we are deeply interested in analyzing
the physics discovered by the ML models: mobility and shear coupling
are dynamic quantities that rely on specific LAEs. Also, since we
don't average the local environment descriptors for LER, including
additional bulk atoms should not affect the outcome of our
descriptor. Indeed, having a larger fraction of the perfect crystal
LAEs in \emph{all} the GBs will not affect how the ML model performs
since the feature is global for all the GBs. Thus, introducing
artificial surfaces is unnecessary for the LER.

\begin{figure*}[htbp]
\centerline{\includegraphics[width=\textwidth]{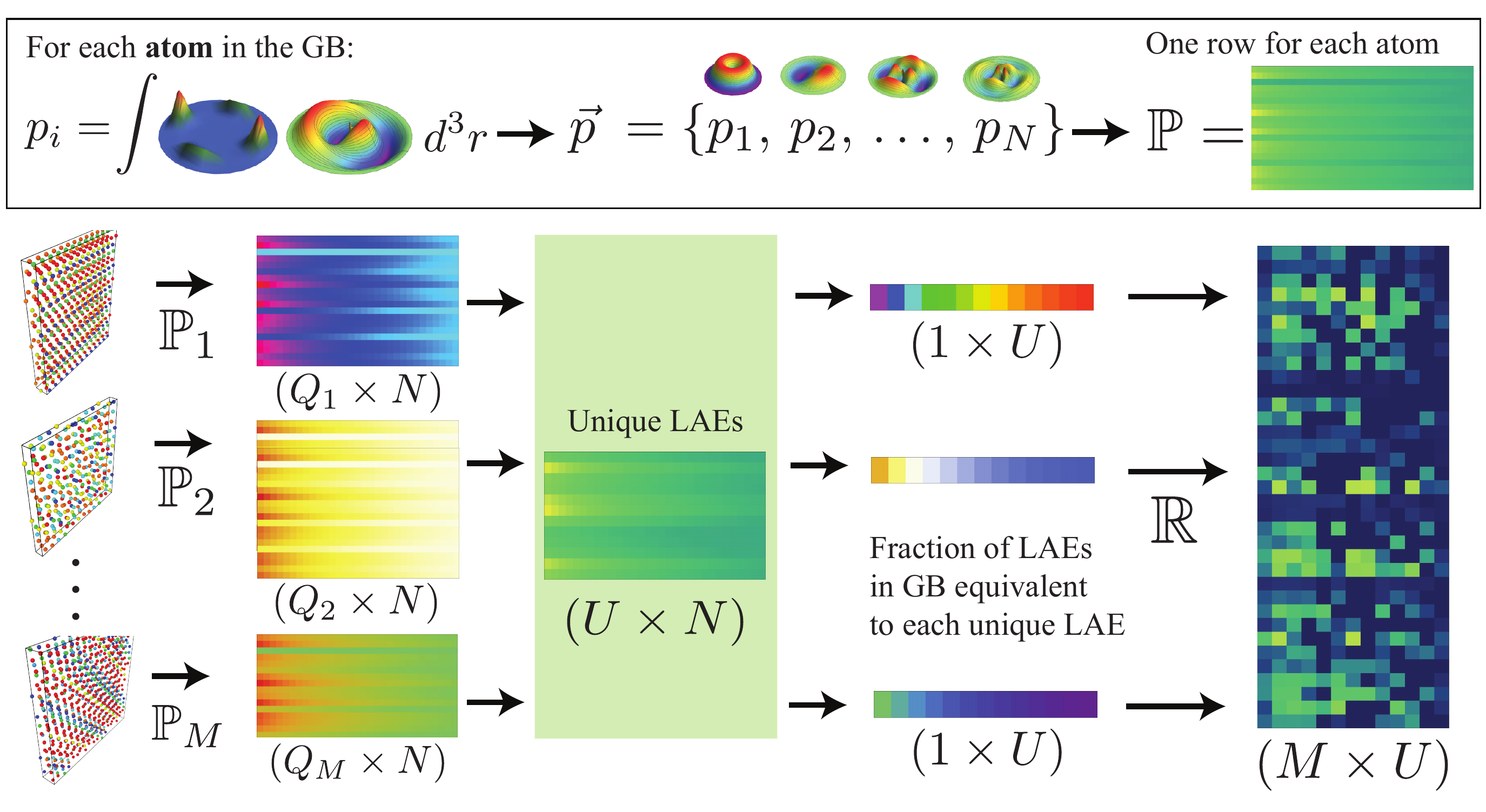}}
\caption[]{\label{fig:gblearnFlowChart} Flowchart for generating the
  SOAP and LER matrices for a single GB and GB system
  respectively. The grain boundary is isolated from the bulk by
  selecting a slab of atoms, as described in Section
  \ref{sec:GBAtomSelection}. Next, each atom in the slab is expanded
  in the SOAP basis to produce a vector of coefficients $\vec{p}$, the
  collection of which forms a matrix $\mathbb{P}$ for each GB in the
  dataset. This matrix can be averaged to form a single vector
  representing the GB, and an ASR matrix for the entire
  system. Alternatively, the similarity metric in the SOAP space can
  be used to find a \emph{unique} subset $U$ of LAEs in the whole GB
  system. Then, for each GB matrix $\mathbb{P}$, a single vector with
  the percentage of unique LAEs can be formed; these are combined to
  form the matrix $\mathbb{R}$, which is the LER.  }
\end{figure*}

Once a collection of GB atoms has been isolated, it is combined into
a \verb|quippy.atoms.Atoms| object from the QUIP library. The
\verb|quippy.descriptors.Descriptor| is initialized using a
string such as \texttt{soap cutoff=5.0 n\_max=18 l\_max=18 \ atom\_sigma=0.5 n\_species=1 species\_Z=\{28\} Z=28 normalise=F}, \\ for the case of
Nickel. For each GB, the descriptor generates a matrix $\mathbb{P}$ where each row is a  $\vec{p}$
(see Eq. \ref{eq:pVectorDefinition}), one for each atom (see Figure
\ref{fig:gblearnFlowChart}).

Through trial and error, we determined that a high bandwidth cutoff
of $l_{\mathrm{max}} = 18$ and $n_{\mathrm{max}} = 18$ was necessary
to get high accuracy for energy predictions. Since forming the LER is
a post-processing step on existing SOAP vectors, we settled on these
values for the free parameters, though a study of some of these parameters is provided in section \ref{sec:paramterstudy}. It is important to maintain
\emph{unnormalized} $\vec{p}$ for all the machine learning
calculations (\verb|normalise=F| in the descriptor string). In every
test we performed with the \emph{normalized} vectors, the performance was
always worse.

\subsection{ASR: GB Atom Selection}
\label{sec:asr_atomselection}

For the ASR \cite{Rosenbrock2017}, we isolated the grain boundary by looking for deviations
from the median values in the centro-symmetry parameter for each atom.
The highest centro-symmetry deviants define the location of the
boundary plane. We then select a rectangular slab 8 \AA\, in width
($\pm 4$ \AA\, to the left and right of the plane) in the direction
perpendicular to the plane of the boundary. A large ($>50$ \AA)
lattice vector in that direction defines a periodic system so that the
local neighborhoods of the new ``surface'' atoms are not affected by
well-ordered bulk.

\subsection{LER: GB Atom Selection}
\label{sec:ler_atomselection}

For the LER \cite{Rosenbrock2017}, we filtered the atoms to find those whose Common Neighbor
Analysis (CNA) value deviated from the perfect bulk (e.g., FCC for
Nickel). The slab is defined by finding the \emph{furthest} non-FCC atoms
in both directions perpendicular to the GB. However, for consistency all LAEs must include a full neighborhood of atoms. Furthermore, since atoms within $ 1 r_{\mathrm{cut}}$ of the slab may include some of these non-FCC atoms, they must be included in the characterization. Therefore, we increase the size of the slab by $ 2 r_{\mathrm{cut}}$ in each direction to ensure they all have a full neighborhood, but only characterize atoms (using $\vec{p}$) within $1 r_{\mathrm{cut}}$ of the slab. This ensures that all non-FCC atoms that contribute to the GB structure are characterized and that LAEs included in the set have a full atomic neighborhood.

\subsection{Tilt GB Characterization: GB Atom Selection}
\label{sec:tilt_atomselection}

In a more recent characterization of a number of $\langle 100 \rangle$ symmetric tilt GBs \cite{Priedeman:2018structure}, we used the same process as the LER noted above (\ref{sec:ler_atomselection}), with a few minor changes. In order to ensure that the characterization was not influenced by the width of any given GB, we used a fixed width for all GBs. This is not unlike the ASR atoms selection method noted above (\ref{sec:asr_atomselection}), except that the fixed width is defined by the widest width of any GB in the set. In this way, GBs will include the same potential width of bulk atoms but ensure that the width is set by GB structure rather than an arbitrary selection of width. 

\subsection{Finding Unique LAEs}
\label{sec:BuildLER}

Once we have the $\mathbb{P}$ matrix for each GB in the system, we can
perform an $n^{2}$ search over all LAEs in the system to find the set
of \emph{unique} LAEs. Because we use the unnormalized $\vec{p}$, a
simple dot product between two arbitrary LAE descriptors may have any
size. We use the following symmetric \emph{dissimilarity} comparison for two
SOAP vectors $\vec{a}$ and $\vec{b}$; it produces a value $d$ that
is 0 for identical LAEs and grows with dissimilarity:

\begin{equation}
\label{eq:UniquifySimilarity}
d = |\frac{||\vec{a}|| + ||\vec{b}||}{2}- \vec{a} \cdot \vec{b}|
\end{equation}

This metric was used in the original method \cite{Rosenbrock2017}, but does not satisfy the triangle inequality. In subsequent analyses \cite{Priedeman:2018structure}, including the analysis of parameters discussed below, we used instead the following dissimilarity metric:

\begin{equation}
\label{eq:UniquifySimilarityNew}
d = \sqrt{\tilde{K}(\vec{a},\vec{a})+\tilde{K}(\vec{b},\vec{b})-2\tilde{K}(\vec{a},\vec{b})},
\end{equation}
which satisfies all the conditions of a dissimilarity metric. This dissimilarity metric can be converted to a similiarity metric via: $s = 1 - d$. 

\begin{figure}[htbp]
\centerline{\includegraphics[width=0.5\textwidth]{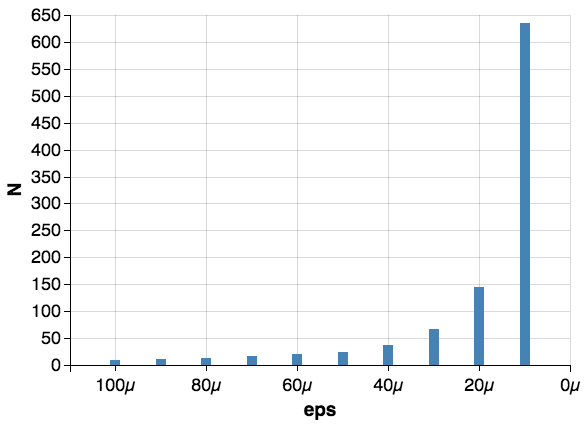}}
\caption[]{\label{fig:epsConvergence} Number of unique LAEs as a
  function of the parameter $\epsilon$. There is a natural cutoff at
  $\epsilon = 20 \mu$ before the number of unique LAEs grows
  exponentially. The reduction from 791K to $\sim150$ is substantial.}
\end{figure}

A finite precision parameter $\epsilon$ controls when two LAEs are
considered identical. In the original application of this method \cite{Rosenbrock2017}, and using the LER atomic selection described above,
we recover a total of 791K atoms and their corresponding LAE. We plot the number of \emph{unique}
LAEs in Figure \ref{fig:epsConvergence} as a function of
$\epsilon$. As epsilon decreases, there is a natural division before
the number of environments grows exponentially. Surprisingly, the
number of unique environments is reduced to approximately $\sim150$ LAEs at
that cutoff.

In the original method \cite{Rosenbrock2017}, we apply the dissimilarity metric (Eq. \ref{eq:UniquifySimilarity}) to the set of LAEs for the purpose of 1) discovering the unique LAEs--as mentioned above-- and 2) assigning a unique LAE ID to each LAE, so as to completely classify the system of LAEs in terms of the unique LAEs of that system. Finally, we
produce a histogram of the unique LAEs for each GB and record the
\emph{fraction} of each LAE to form an LER vector for the GB. The
collection of these forms the LER matrix (see Figure
\ref{fig:gblearnFlowChart}).

More recently \cite{Priedeman:2018structure}, we refined the selection of the unique LAEs. In \cite{Rosenbrock2017}, the discovery of unique LAEs and the assignment of a unique LAE identifier to each LAE occurred in the same step. This old methodology relied on an incomplete set of unique LAEs to classify the LAEs (since discovery and assignment occurred concurrently). The method also used the first unique LAE satisfying the equivalence threshold to classify a given LAE. Therefore, the methodology was modified in \cite{Priedeman:2018structure} to address these issues. The first change divides unique LAE discovery and assignment into separate steps. For the discovery step, the new method iterates through the set of LAEs to discover the unique LAEs, without making any assignments. For the assignment step, the method re-iterates through the set of LAEs, to assign each LAE a unique LAE identifier (from the complete list of unique LAEs). The second change lies at the core of the assignment process: rather than assigning the first unique LAE that satisfies equivalency to represent an LAE, the similarity of each unique LAE to the LAE in question is evaluated. The unique LAE with the greatest similarity to the considered LAE (equivalence is still a requirement) is assigned to represent the considered LAE. This results in disjoint unique LAE sets. This modified methodology is used to execute the parameter study, detailed in Section \ref{sec:paramterstudy}. 

In the recent characterization of $[1\ 0\ 0]$ - symmetric tilt GBs \cite{Priedeman:2018structure}, we began by analyzing the 388 GBs created by Olmsted \cite{Olmsted2009}, along with 100 new $[1\ 0\ 0]$ - symmetric tilt GBs simulated at BYU by Erickson \cite{Erickson} and Priedeman \cite{Priedeman:2018structure} to determine the number of atoms in the GBs and the number of unique LAEs (using the modified method). We use the parameters determined in Section \ref{sec:paramterstudy} for this analysis. In total, there are 1053K atoms for this set of 488 GBs, with 66 unique LAEs. If we use the same parameters but exclude the 100  $[1\ 0\ 0]$ - symmetric tilt GBs, analyzing only Olmsted's 388 GBs, there are 590K atoms and 62 unique LAEs. 

\section{Influence of SOAP parameters and equivalence threshold on LER}
\label{sec:paramterstudy}

    As indicated by the free variables noted above, the SOAP descriptor requires a set of input parameters to produce a descriptor of a given structure. These parameters are critical to capturing the necessary information about the environment, as poor parameter selection will cause the descriptor to capture at best, unimportant, or at worst, distracting, information. Here we present an examination of the following parameters that influence the SOAP and LER characterizations: 1) the cutoff radius $r_{cut}$, 2) the number of angular basis functions in the spherical harmonics $l_{max}$, 3) the number of radial basis functions $n_{max}$, 4) the Gaussian width $\sigma$, and 5) the equivalence threshold $\epsilon$ (which is a parameter for the local environment representation). The evaluation of each metric is focused on how it influences the number of unique LAEs produced. This parameter study was undertaken in anticipation of the work by Priedeman et al. \cite{Priedeman:2018structure}, which followed the work of Rosenbrock et al. \cite{Rosenbrock2017} .
	
\subsection{Cutoff Radius}

\begin{figure*}[htbp]
	\centering
	\includegraphics{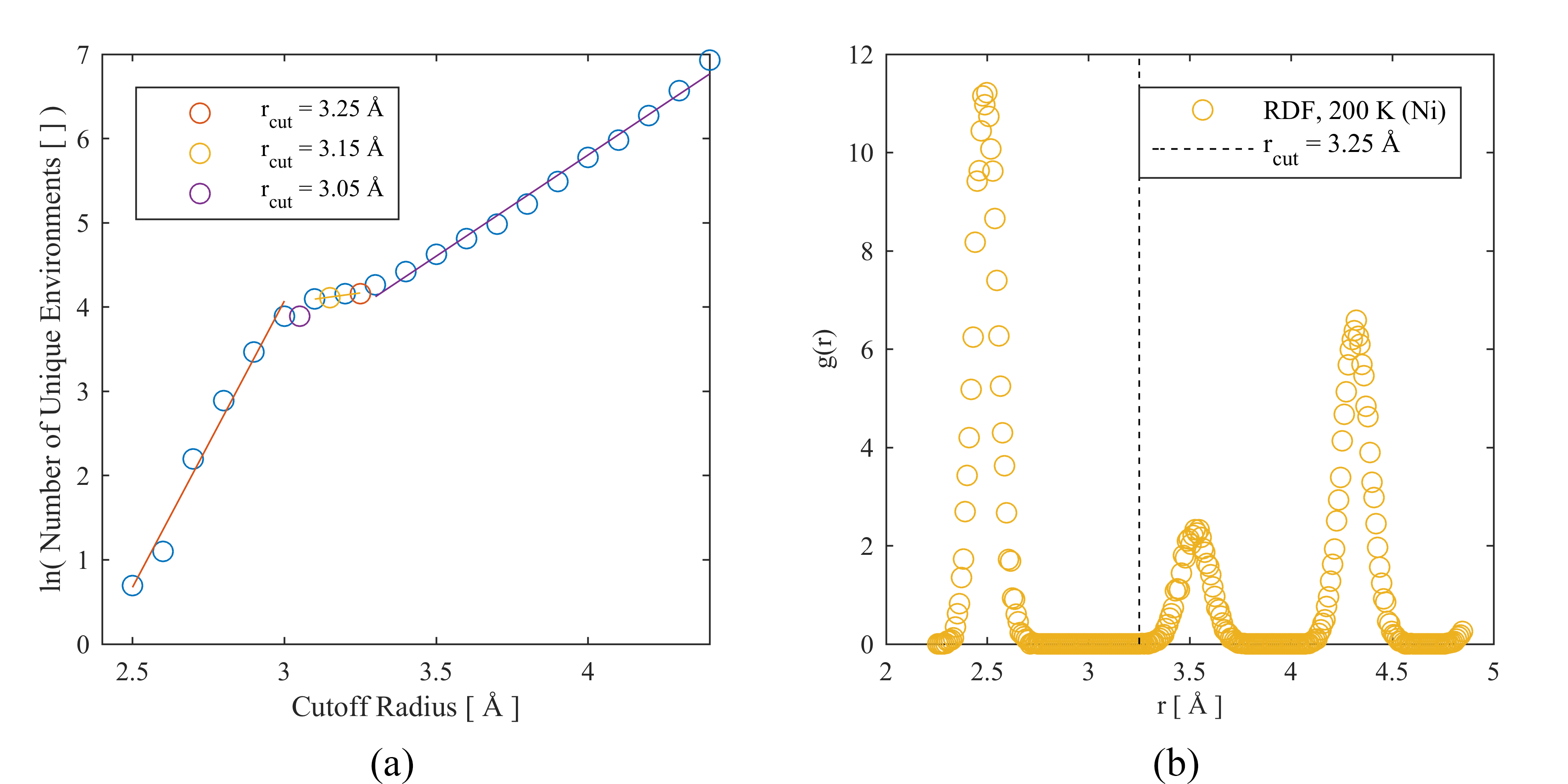}
	\caption[Cutoff Radius ($\protect{r_{cut}}$) Parameter Selection]{(a) Variation of the number of unique LAEs as a function of the cutoff radius ($r_{cut}$). (b) The radial distribution function in FCC nickel at 200 K, with the chosen cutoff radius marked by a dashed line. The selection of the cutoff radius allows for the SOAP descriptor to capture information almost out to the second ``ring" of nearest neighboring atoms (the second spike in the RDF).}
	\label{fig:ULAE_Rcut}
\end{figure*}

    The cutoff radius $r_{cut}$ defines the extent of the environment being described. A SOAP description using a small value for the cutoff radius will only describe local atomic structure, while a description using a large cutoff radius value will capture information about the atomic structure of a larger region. To investigate the cutoff radius, we apply the SOAP descriptor to the atomic structures of the 388 GBs from Olmsted's dataset \cite{Olmsted2009}, using a range of cutoff radii. With these SOAP descriptions, we determine the number of unique LAEs produced for each cutoff radius. For this analysis, the other SOAP parameters are: $l_{max}$ = 18, $n_{max}$ = 18, $\sigma$ = 0.5 \AA. We plot the number of unique LAEs against the cutoff radius ($r_{cut}$) used in the analysis in Fig. \ref{fig:ULAE_Rcut}(a). As expected, the number of unique LAEs increases as $r_{cut}$ increases, since there are more ways to uniquely position the atoms in the local environment. However, in the range of 3.05 \AA\  $<$ $r_{cut}$ $<$ 3.25 \AA, the number of unique LAEs is almost constant. We identify this brief plateau as the ideal range for $r_{cut}$, since the number of unique LAEs appears to be independent of $r_{cut}$; we select the upper end of this range ($r_{cut}$ = 3.25 \AA) to be as inclusive as possible. For the investigation presented in \cite{Priedeman:2018structure}, visual comparison of the unique LAE classification of the C-structural units for $r_{cut}$ = 3.05, 3.15, and 3.25 \AA\ indicated that $r_{cut}$ = 3.25 \AA\ yielded the greatest consistency in the classification of these C-structural units. 

	There is some physical meaning to a cutoff radius of 3.25 \AA. The lattice parameter a\textsubscript{0} has a value in nickel of a\textsubscript{0} = 3.52 \AA \, at 0 K, with the first nearest neighbors at 2.49 \AA \, and the second nearest neighbors at 3.52 \AA. At $r_{cut}$ = 3.25 \AA, we capture atom positions that would almost be in the second nearest neighbors (in perfect FCC). We illustrate this point with Fig. \ref{fig:ULAE_Rcut}(b), which depicts the radial distribution function for FCC nickel at 200 K. 

\subsection{Number of Basis Functions}

\begin{figure*}[htpb]
	\centering
	\includegraphics{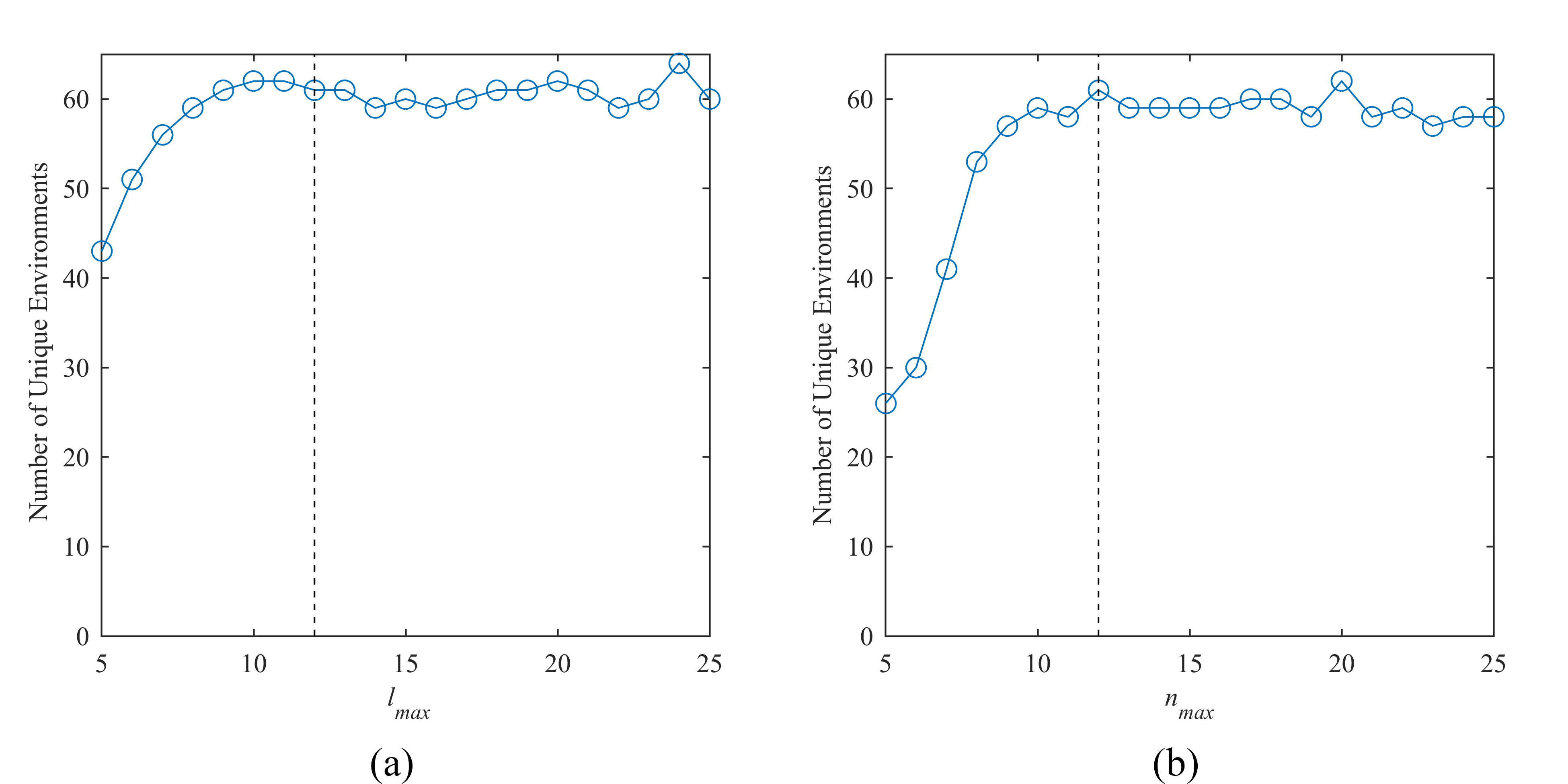}
	\caption[Number of Basis Functions ($\protect{l_{max}}$, $\protect{n_{max}}$) Parameter Selection]{(a) Plot of the number of unique LAEs as a function of the number of angular basis functions in the spherical harmonics ($l_{max}$). The dashed line indicates the selected value. (b) Plot of the number of unique LAEs as a function of the number of radial basis functions in the spherical harmonics ($n_{max}$). The dashed line indicates the selected value.}
	\label{fig:ULAE_lnmax}
\end{figure*}

    The number of basis functions (in the SOAP descriptor) plays a similar role to the number of sinusoidal terms in a Fourier transform: a greater number of functions or terms will increase the fidelity of the description. To investigate the influence of the number of angular basis functions in the spherical harmonics on the SOAP descriptor and the local environment representation, we again describe the atomic structures of the 388 GBs from Olmsted's data set \cite{Olmsted2009}, this time varying the number of angular basis functions in the spherical harmonics ($l_{max}$) rather than the cutoff radius. For this description, the other SOAP parameter values are: $r_{cut}$ = 3.25 \AA, $n_{max}$ = 9, $\sigma$ = 0.5 \AA. The SOAP descriptions are analyzed to determine the number of unique LAEs. We expect to see the emergence of an asymptotic trend between the number of unique LAEs and the number of angular basis functions: once the dominant angular basis functions are included in the SOAP description, additional angular basis functions will only slightly modify the description. Fig. \ref{fig:ULAE_lnmax}(a) plots the number of unique (local atomic) environments against the number of angular basis functions in the spherical harmonics ($l_{max}$). We see that the asymptotic trend is present, as expected. We choose our number of angular basis functions ($l_{max}$) to be 12 (vertical dashed line in Fig. \ref{fig:ULAE_lnmax}(a)) so as to fall within the asymptotic regime without incurring too much computational cost. 
	
	To investigate the influence of the number of radial basis functions ($n_{max}$), we repeat the analysis above, varying the number of radial basis functions ($n_{max}$) instead. The other SOAP parameter values are: $r_{cut}$ = 3.25 \AA, $l_{max}$ = 9, $\sigma$ = 0.5 \AA. Another asymptotic trend is expected between the number of unique LAEs and the number of radial basis functions. Fig. \ref{fig:ULAE_lnmax}(b) plots the number of unique (local atomic) environments against the number of radial basis functions ($l_{max}$), and confirms our intuition regarding an asymptotic trend. To fall within the asymptotic behavior while incurring minimal computational cost, we select $n_{max}$ = 12. 
	
\subsection{Gaussian Width}
	
\begin{figure}[tbp]
	\centering
	\includegraphics[width=\columnwidth]{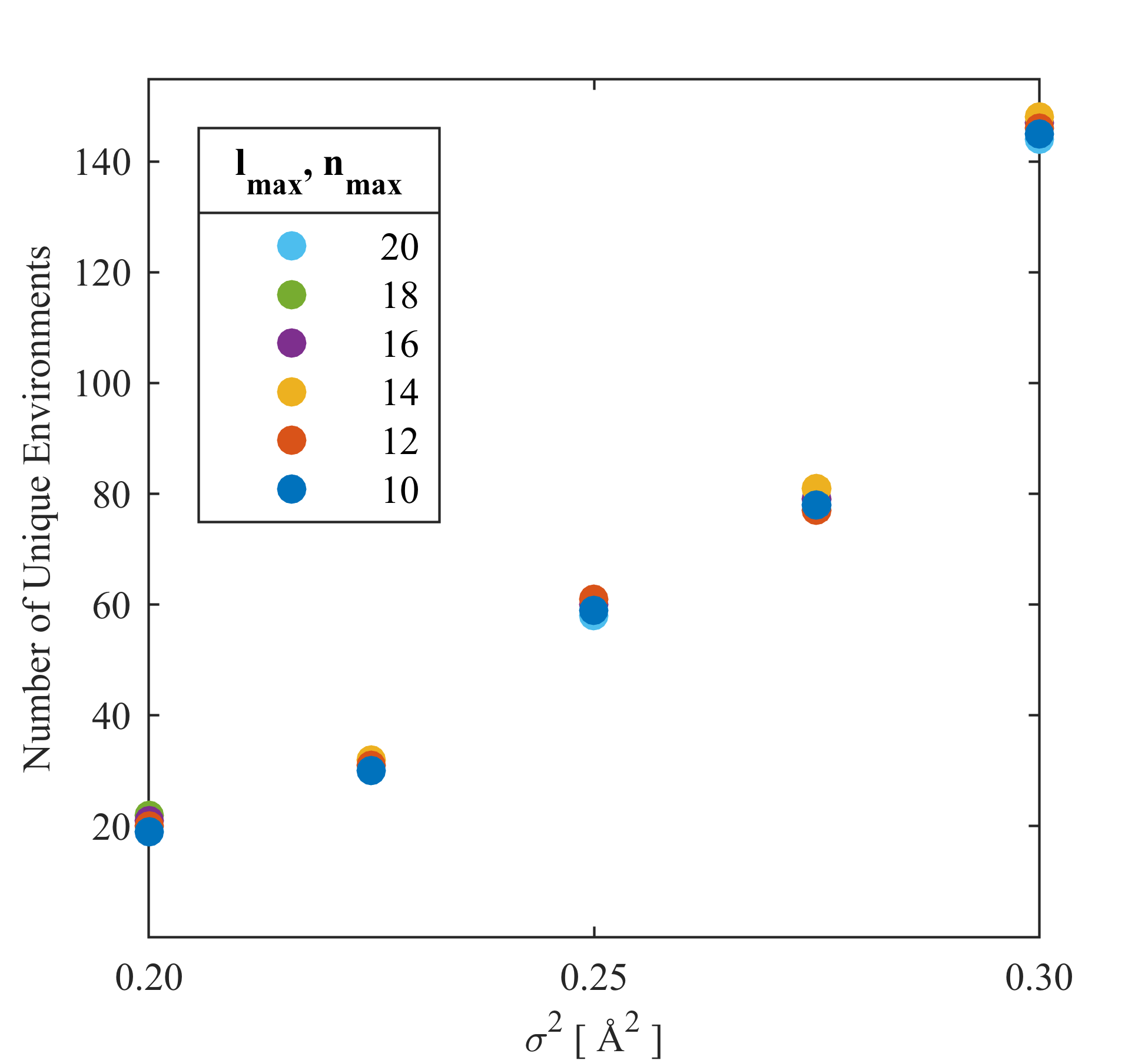}
	\caption[Gaussian Width ($\protect{\sigma}$) Parameter Selection]{Plot of the number of unique LAEs as a function of the Gaussian width, for several different $l_{max}$, $n_{max}$ values.}
	\label{fig:ULAE_LNS}
\end{figure}

	The Gaussian width ($\sigma$) determines the extent of the three-dimensional densities placed at each atom's location, in other words, it serves as the translation between a distribution of points (the actual atom locations) and a density distribution (which can be more easily captured by smoothly varying functions). The Gaussian width influences the smoothness of the descriptor \cite{Deringer2017}, and so we also vary the number of basis functions (treating $l_{max}$ and $n_{max}$ as a single, linked variable) to attempt to observe any changes in the number of unique LAEs produced. The cutoff radius is fixed at 3.25 \AA. The results are plotted in Fig. \ref{fig:ULAE_LNS}. Since there was not a compelling argument to changing the $\sigma$ value, it was left at $\sigma$ = 0.5 \AA\ ($\sigma^2$ = 0.25 $\AA^2$) . 
	
\subsection{Equivalence Threshold}

\begin{figure}[tbp]
	\centering
	\includegraphics[width=\columnwidth]{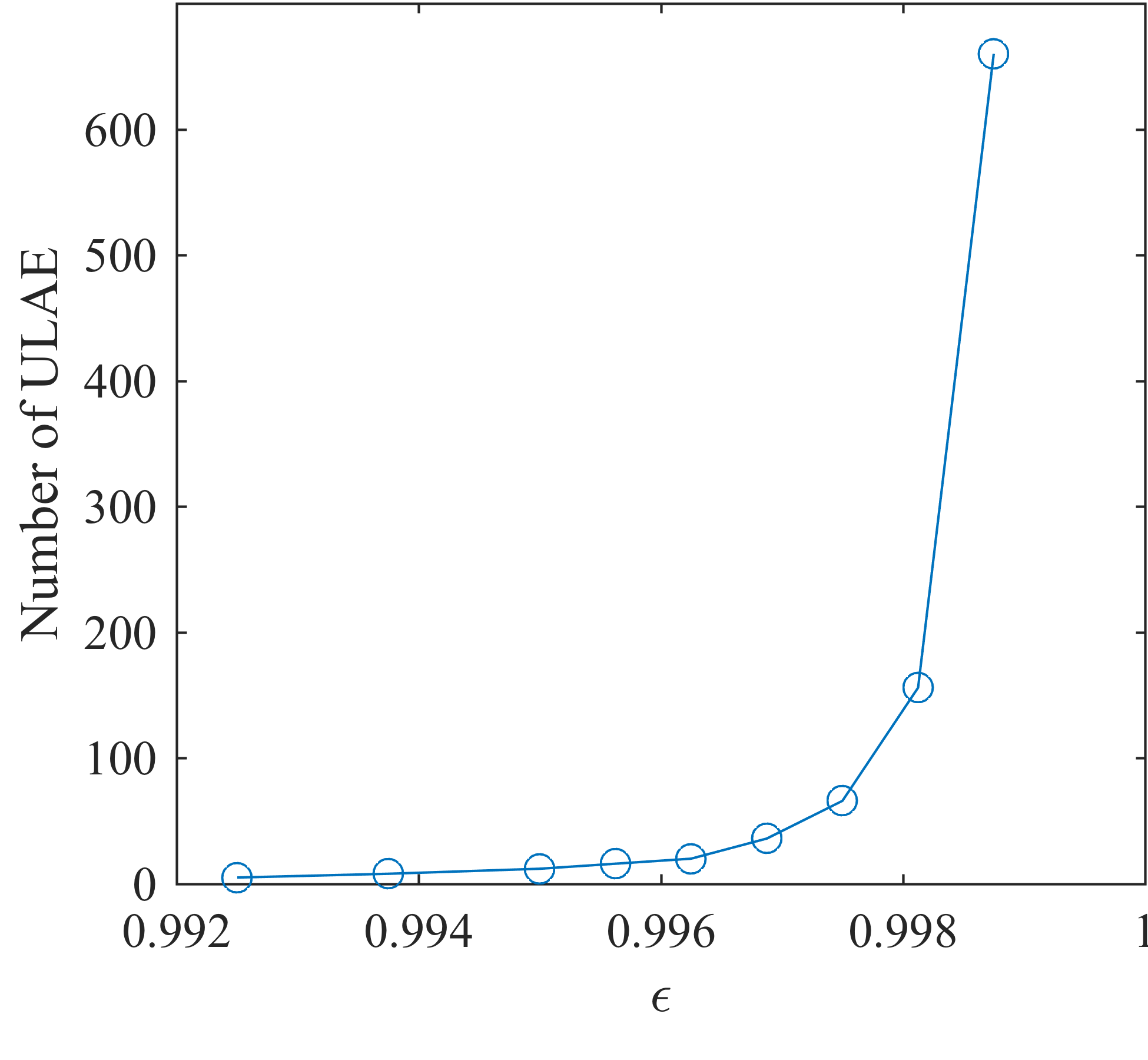}
	\caption[Equivalence threshold ($\protect{\epsilon}$) Parameter Selection]{Plot of the number of unique local atomic environments as a function of the equivalence threshold. The trend appears to be exponential, so we arbitrarily chose a value at a point before the number of environments begins to increase dramatically. Note that the dissimilarity metric (Eq. \ref{eq:UniquifySimilarityNew}) is converted to a similarity metric here: $s = 1 - d$. }
	\label{fig:ULAE_eps}
\end{figure}

	The equivalence threshold ($\epsilon$) is not a parameter of the SOAP descriptor, but of the local environment representation. The equivalence threshold delimits when two LAEs would be considered equivalent (essentially describing identical environments). We examine the number of unique local atomic environments associated with different equivalence thresholds. The SOAP parameters are fixed at values of: $r_{cut}$ = 3.25 \AA, $l_{max}$ = $n_{max}$ = 12, and $\sigma$ = 0.5 \AA. Fig. \ref{fig:ULAE_eps} depicts the number of unique (local atomic) environments against the equivalence threshold $\epsilon$, and we observe the trend to appear to be exponential. We ultimately chose $\epsilon$ = 0.9975 for an investigation of the connection between GB crystallography and atomic structure \cite{Priedeman:2018structure}; at this value, the exponential trend had not yet increased dramatically, still included several dozen unique LAEs, and produced consistent characterization of known structures, like the C-structural unit.
	
\subsection{Parameter study conclusions}
	
	This parameter study was not meant to provide an exhaustive examination of each parameter's influence on the GB structural characterization or to determine what makes for an ideal characterization. Rather, after the original work  was completed \cite{Rosenbrock2017}, we realized that we might get better characterization if we used different parameters. These new parameters have the following benefits (compared to the previously used parameters in \cite{Rosenbrock2017}):
\begin{itemize}
	\item Consistent patterns in the local environment representation of the C-structural unit (from the structural unit model) \cite{Priedeman:2018structure}. 
	
	\item A smaller set of unique LAEs that still has good prediction capability. Rosenbrock repeated his machine learning (described in \cite{Rosenbrock2017} and below) using the revised SOAP parameters and saw slightly better prediction of GB properties in the repeated simulation over the original. This slight improvement comes even as the machine uses less than half the number of unique local atomic environments and one third the number of coefficients in the SOAP vectors (1,015 instead of $\sim$3,000) compared to the original input.
\end{itemize}
	
	Obviously, a more robust and thorough examination must be executed to determine if there are ideal values for GB structural characterization and what those values should be.  However, before investing significant time determining what ideal values should be used, we determined that it was worth demonstrating what information can be gained using a set of briefly selected optimal values.
	
	We recognize that of the parameters that we have examined so far, the equivalence threshold selection is the one that we judge to be most arbitrary, with the Gaussian width a close second. This is because 1) the equivalence threshold is the hardest parameter to conceptualize (comparing fits to three-dimensional densities, even though it is similar to a dot product) and 2) we currently have no benchmarks for recognizable differences between environments. (For example, at what value of $\epsilon$ would two defects known to be unique, be classified as identical?)

\section{Machine Learning}

We now demonstrate the manner in which the ASR and LER are used in
the machine learning of GB energies and GB mobility and shear coupling.


\subsection{Grain Boundary Energy}

Grain boundary energy was learned using the kernel similarity matrix
derived from the ASR (see Eq. (\ref{eq:FinalTildeK})). Although the
ASR defined above used a simple vector averaging of the SOAP
$\mathbb{P}$ matrix, it is also possible to average ``globally''
(\verb|average=T| in the descriptor string) so that all the atoms are
treated equivalently to produce a single descriptor for the
\emph{collection} of atoms. This descriptor is quite different from
averaging individual atomic descriptors afterwards; in fact, at first
glance, averaging globally is the natural choice because it preserves
the relative positioning of each atomic environment to all the others
in the final, averaged descriptor. Averaging the LAE descriptors
externally loses that information. In our experiments, however, we
concluded that averaging externally was twice as accurate as averaging
globally, which was unexpected.

\begin{figure}[tbp]
\centerline{\includegraphics[scale=0.5]{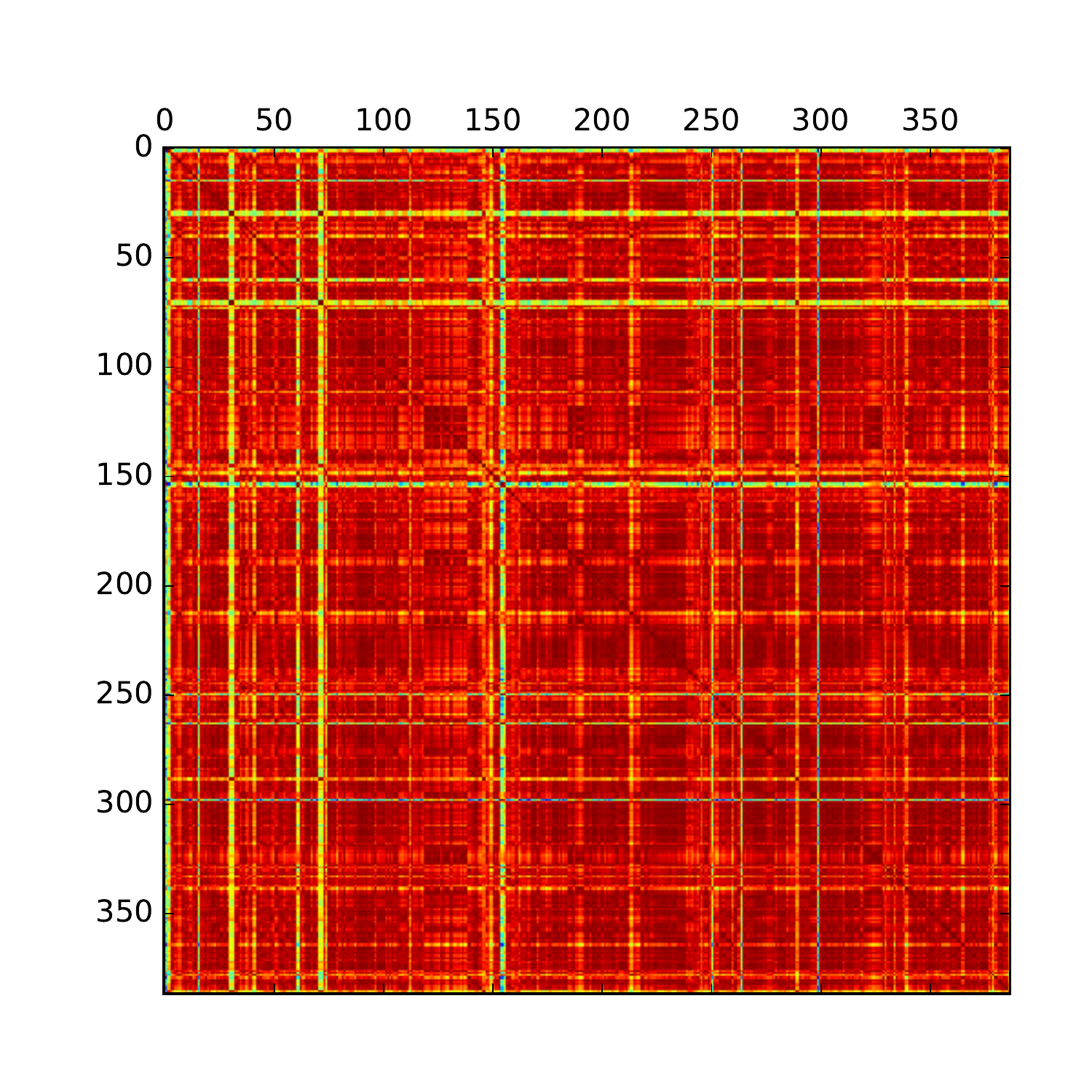}}
\caption[]{\label{fig:KMatrix} A matrix plot of the kernel matrix $K$
  between the structures in the Olmsted data set for Nickel (from
  Eq. (\ref{eq:FinalTildeK})). A random subset of this matrix was used
  to train the SVM.}
\end{figure}

To test the quality of the new descriptor, we applied it to the
calculation of grain boundary energies using machine learning. We used
the structures and energies of Olmsted \cite{olmsted2009survey} and
calculated the SOAP descriptor $p$ for each structure to form the
similarity matrix $K$, plotted in Figure \ref{fig:KMatrix}. For the
SOAP descriptor, we used a neighbor cutoff $r_{\mathrm{cut}}=5.0$ \AA,
and expansion cutoffs of $n_{\mathrm{max}}=18$ and
$l_{\mathrm{max}}=18$ (``soap cutoff=5.0 n\_max=18 l\_max=18
n\_species=1 species\_Z={28} Z=28 normalise=F'' is the corresponding
descriptor string).

\begin{figure}[tbp]
\centerline{\includegraphics[scale=0.45]{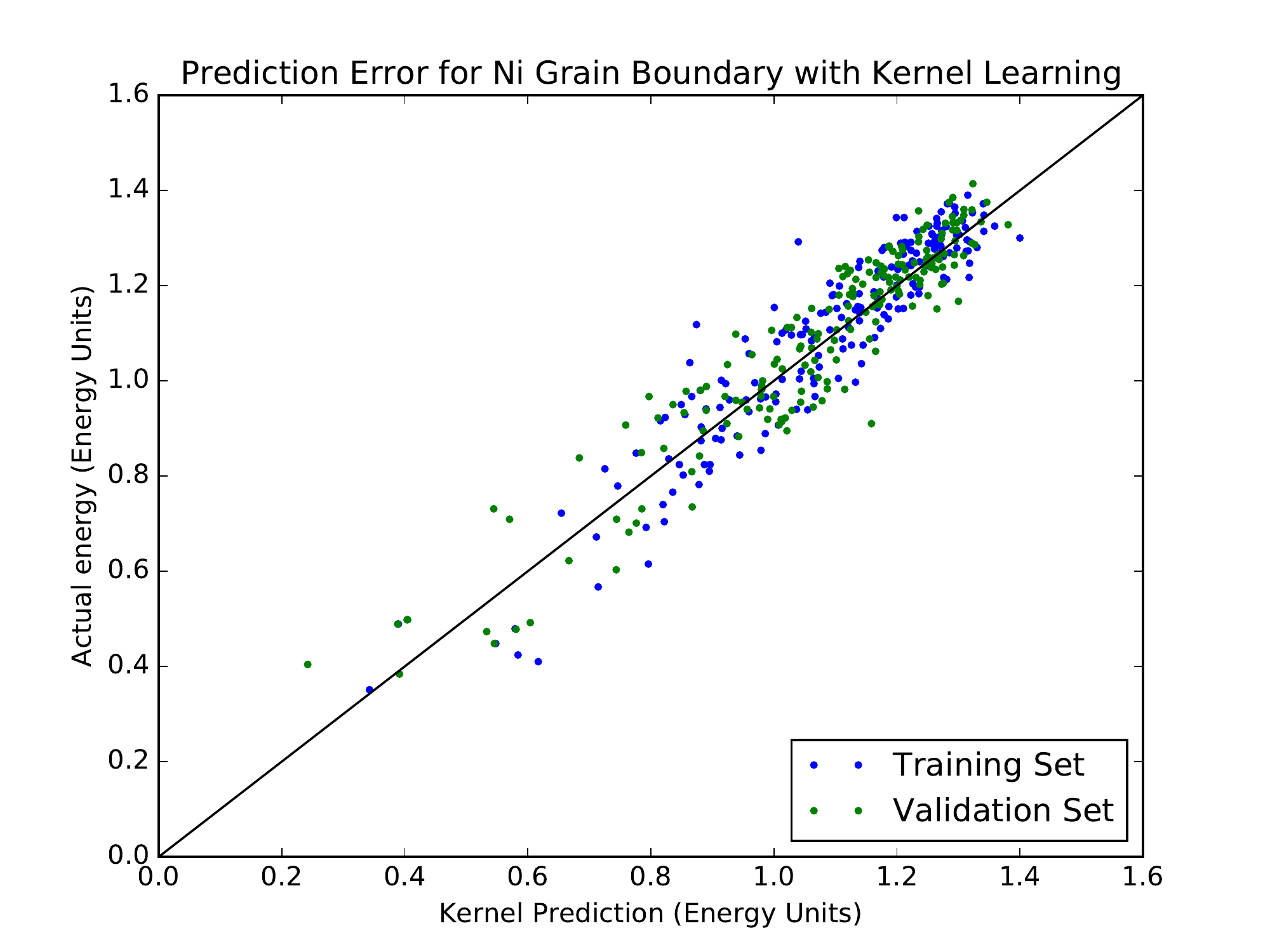}}
\caption[]{\label{fig:NiPrediction} The SVM energy predictions for Ni
  in the Olmsted data set, trained using 50\% of the data; the
  remaining 194 data points were used to validate the model. Although
  the RMS devation from the exact answer is appreciable, the
  predictions are definitely correlated. Using more data points will
  likely narrow the predictions. }
\end{figure}

A random subset of 50\% of the available data ponts (194 structures)
defined the training set for the SVM. We used the Support Vector
Regression implementation in \textsc{sklearn} \cite{scikit-learn},
based on \textsc{libsvm} \cite{CC01a}. The $C$ and $\gamma$ parameters
were found via logarithmic grid search in the ranges [$10^0$, $10^3$]
and [$10^{-2}$, $10^2$] respectively to be $C=1000$ and
$\gamma=100$. After training the SVM, we validated the model using the
remaining data points as shown in Figure \ref{fig:NiPrediction}.

The RMS error (0.07) is not excellent (the standard deviation in the
energies is 0.37). However, the size of the data set is small by
machine-learning standards. If we were to include several thousand
more examples to learn from, the SVM's predictions would likely
improve significantly.

\subsection{Grain boundary mobility and shear coupling}

As an additional illustration of the usefulness of the descriptor, we
also trained a ML model to classify the mobility and shear-coupling
types of each grain. For grain boundary mobility, we divided the GBs
into four classes: 1) Immobile (I); 2) Thermally Activated (TA); 3)
Thermally Damped (TD); 4) Constant(C). For GB shear coupling, we
classified over two coupling types: 1) Not Coupled (NC); 2) Shear
Coupled (SC). The data for these ML exercises comes from previous publications involving the Olmsted dataset of 388 GBs for mobility \cite{Homer:2014hr} and shear coupling \cite{Homer:2013ce}.

\begin{figure}[tbp]
\centerline{\includegraphics[scale=0.45]{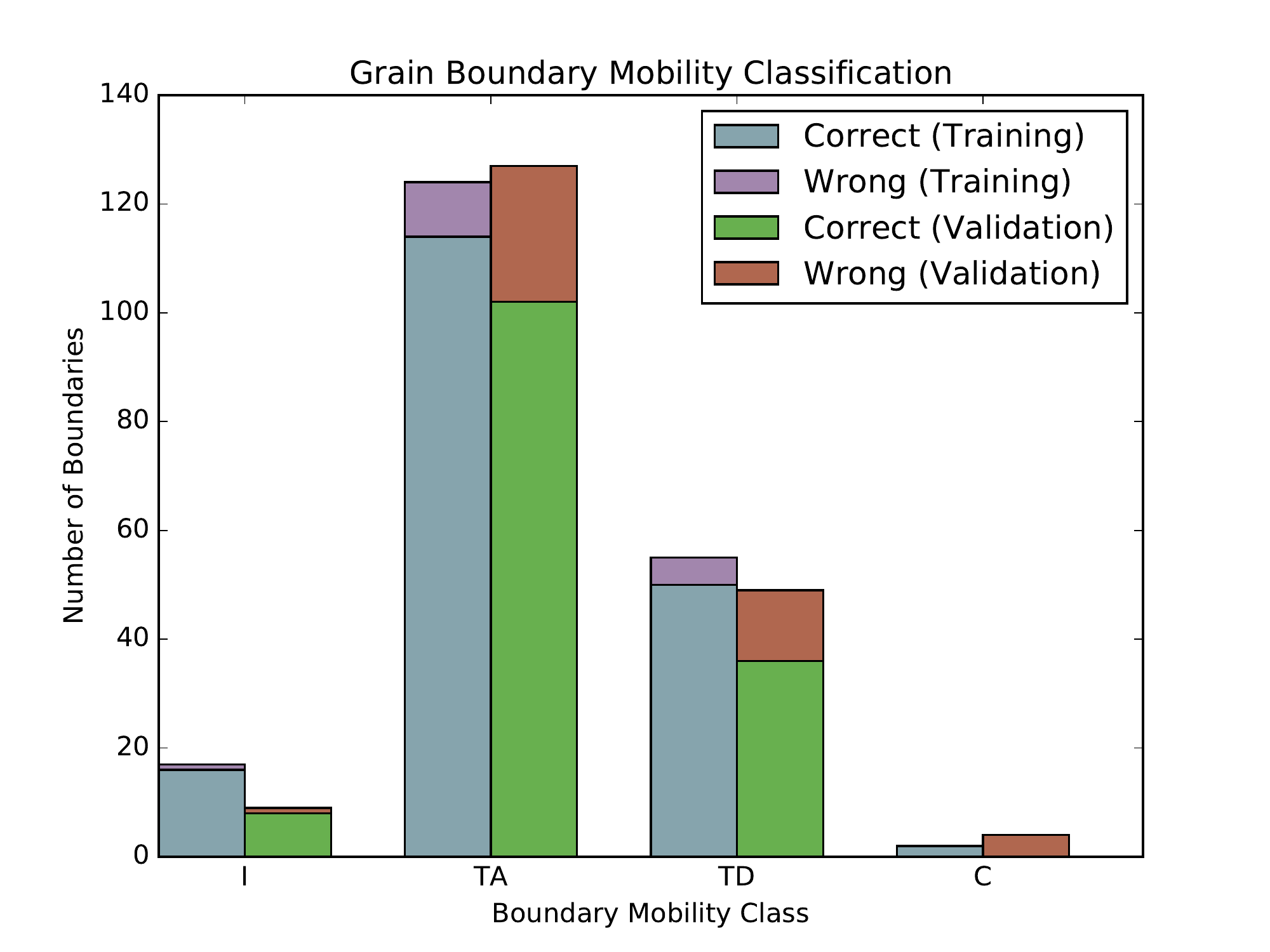}}
\caption[]{\label{fig:GBMobilityPrediction} Prediction of grain
  boundary mobility type using SVM classification. The SVM predicted
  the correct class with 91.9\% and 77.2\% accuracy for the training
  and validation sets repsectively. SVM predictions for the constant
  class (C) are terrible because of a lack of training data in that
  class.}
\end{figure}

We first attempted to use a SVM approach to classification using the
ASR (\emph{not} the kernel matrix $\tilde{K}$). For each test, we
trained a linear support vector classifier model using an $\ell_1$
norm penalty function with parameter $C$ optimized via grid-search
over a logspace from [10$^{-2}$, 10$^0$]. The machine trained on
approximately half of the available data points with the prediction
being verified on the other half. Results for the mobility
classification are plotted in Figure
\ref{fig:GBMobilityPrediction}. Generally, the SVM classifier does
well and is able to predict mobility with 77\% accuracy on the
validation set. Once again, these results use out-of-the-box
algorithms and procedures with no fine tuning or by-hand
preconditioning of the feature space. This result is significant
because it proves that 1) the mobility of the grain boundaries is a
function of the local environment at the boundary; 2) the aggregated
SOAP descriptor captures a reasonable amount of the essential atomic
environment information to enable GB mobility prediction.

\begin{figure}[tbp]
\centerline{\includegraphics[scale=0.45]{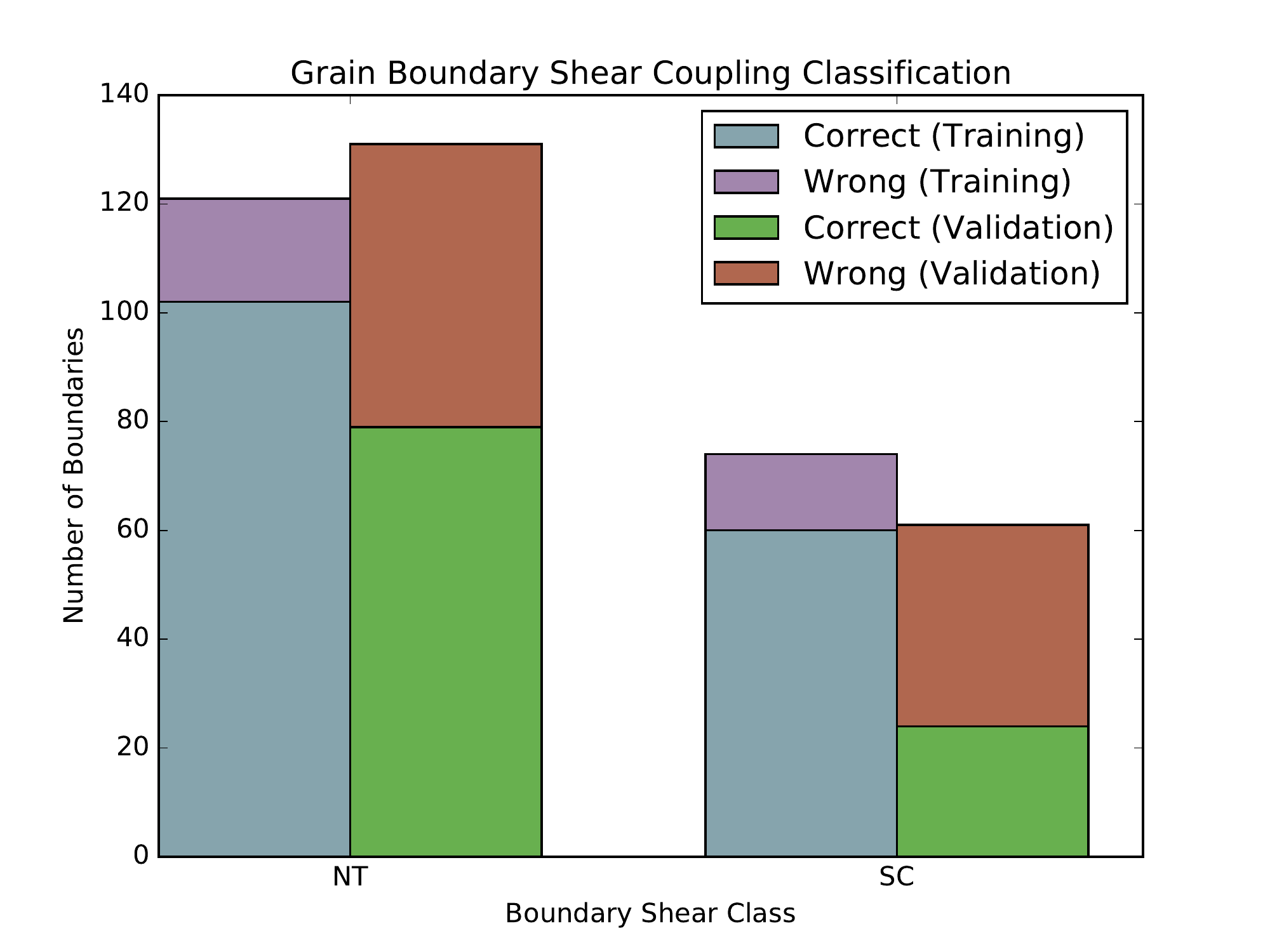}}
\caption[]{\label{fig:ShearCouplingPrediction} Prediction on whether a
grain boundary is shear coupled or not. The classifier was 83.0\% and
53.7\% accurate on the test and training sets respectively. This
prediction is slightly better than random and provides no physical
insight into physical mechanisms since it uses the ASR.}
\end{figure}

The results for the shear coupling classification are shown in Figure
\ref{fig:ShearCouplingPrediction}. At first glance, they are
disappointing since the classifier can scarcely do better than
random. Additionally, neither the mobility nor the shear coupling
results offer any physical insight into the local mechanisms for these
phenomena. The support vectors live in the same space as the SOAP
vectors from the ASR; they are used to construct the hyperplane that
divides different classes. While they can be used to map out the
hyper-dimensional surfaces, the high dimensionality of the result
makes it hard to interpret. This motivates us to use a different
representation that could do as well (or better) in classification
accuracy, but \emph{also} offer physical interpretability. The LER
provides the solution.

The methods for constructing the LER were discussed in Section
\ref{sec:BuildLER}. Once constructed, we use the \verb|XGBClassifier|
from the excellent \verb|xgboost| library
\cite{Chen:2016:XST:2939672.2939785} to produce a classification
model. \verb|xgboost| uses a collection of gradient boosted decision
trees (GBDTs) to iteratively improve the model at its weakest
points. Although we did perform a grid search over all recommended
parameters (requires fitting 25920 separate models with 5-fold
cross-validation), we determined that the default parameters for the
model perform essentially as well as the best-tuned models.

Because of the imbalanced class problem in the mobility
classification, we attempted a variety of standard methods using the
\verb|imbalanced-learn| library \cite{lemaitre2016imbalanced}. The
borderline SMOTE method
\cite{Han:2005:BNO:2141202.2141297,Nguyen:2011:BOI:1972030.1972031}
provided the best performance. By oversampling with SMOTE and using
the \verb|XGBClassifier|, we were able to push the performance on
mobility classification beyond 80\% accuracy. There are two different
approaches for learning a multi-classification problem: 1) train a
machine to predict all the classes simultaneously; 2) train a series
of machines to predict binary combinations of classes. We found that
the machine trained on the multi-class problem \emph{always}
out-performed the binary classifiers. The multi-class classifier
achieved 85.5\% accuracy on the validation set when the constant (C)
class was ignored. We chose to ignore that class because it is
severely under-represented. A dataset of only 388 samples is already
small by ML standards; attempting to learn which LAEs contribute to
constant mobility from only ~5 GBs is apparently not possible using this representation. When more
data becomes available we are confident that the current framework
will successfully learn that class as well.

\begin{figure}[tbp]
\centerline{\includegraphics[scale=0.45]{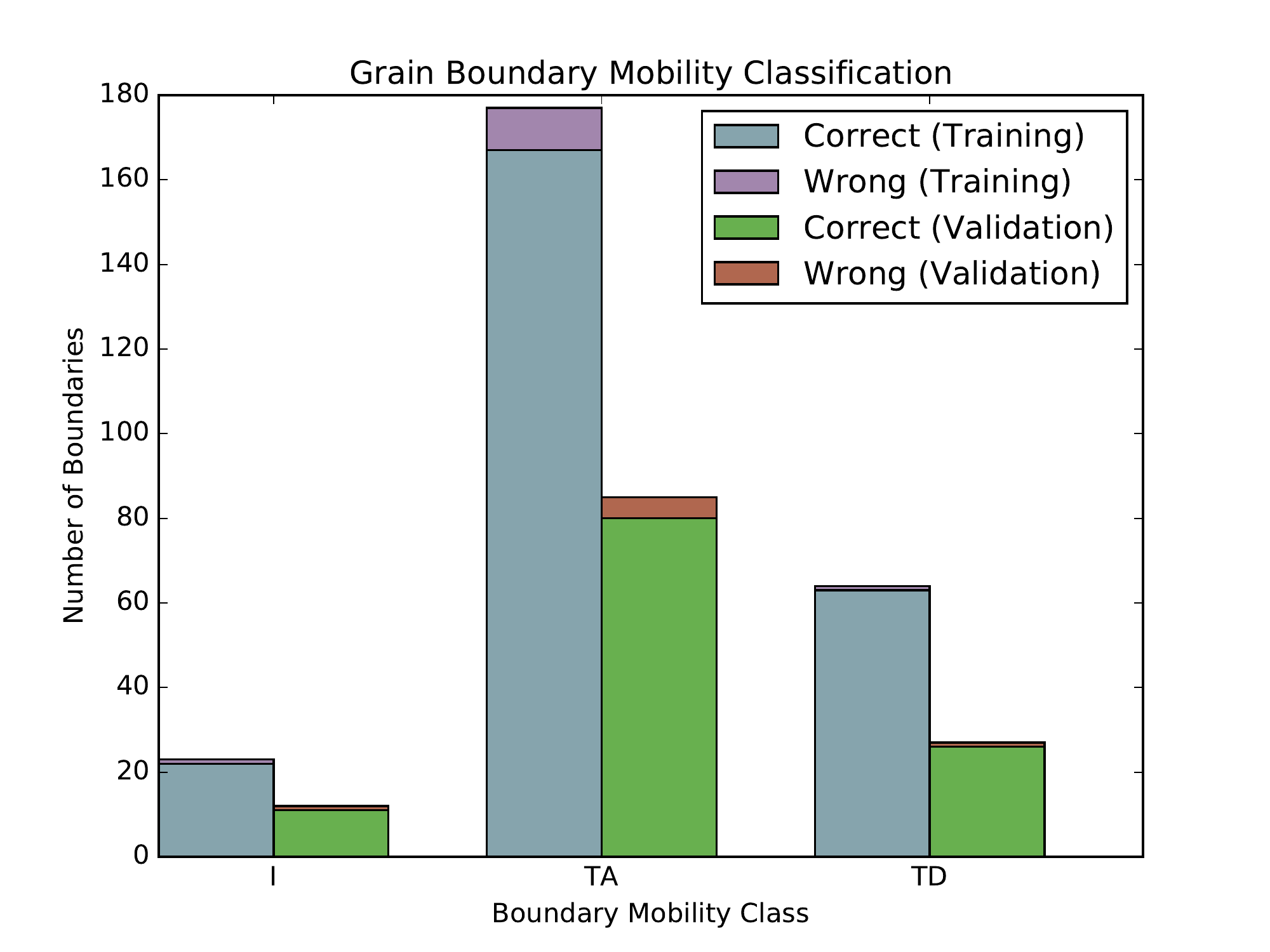}}
\caption[]{\label{fig:MobilityXGBT1} Classification of the mobility
  type for grain boundaries using the LER. The classes were first
  re-sampled using borderline SMOTE, and then 66\% of the GBs were used
as a training set. The constant class was omitted from the problem
because it has too few representatives to be learnable.}
\end{figure}

For shear coupling, we were able to increase prediction accuracy to
64\%, though the solution used a linear SVM with $\ell_1$
regularization. Because the decision tree classification did not
perform as well, we cannot interpret the origins of shear coupling in
the LAE framework as we can with mobility. This may indicate that
shear coupling prediction requires more information than merely a
knowledge of local environments.

\section{Concluding remarks}

Creating useful structure-property relationships is challenging because of the many ways GBs can be structured. The five degrees of freedom for the crystallographic structure is a large space of possibilities that is not well explored. The 3$N$ dimensional space of atomic structure, for $N$ atoms in a grain boundary, is even larger and less well explored. A systematic discovery of these spaces is formidable and not likely to be accomplished very soon. But, careful characterization of the limited data we presently have available, combined with state of the art machine learning can provide powerful insight.

In this work we detail a recent atomic descriptor of GBs, which has advantages over other atomic descriptors \cite{Priedeman:2018structure}. This new descriptor is beneficial because it provides a smooth, differentiable metric that can be used for comparison of different structures. The new descirptor makes use of the SOAP local
environment descriptor. It has all the necessary physical invariances
and the smoothness and differentiability that make it especially well
suited to machine learning problems. 

To show the usefulness of the new descriptor, we
apply it to energy prediction using support vector regression, and
to mobility and shear coupling classification using support vector
classification and gradient-boosted decision trees. The results are encouraging and suggest that some structurally universal building blocks of grain boundaries may exist.

The inclusion of more data points will likely generate a
model that can quickly predict grain boundary energies for any
single-species grain boundary with sufficient accuracy to be
useful. This will enable faster determination of the most stable grain
boundary structures across a larger configurational space.

\section*{Acknowledgments}
The authors thank G\'{a}bor Cs\'{a}nyi for his help in applying SOAP to grain boundaries in the original method. C.W.R. and G.L.W.H. are supported under ONR (MURI N00014-13-1-0635). J.L.P. and E.R.H. are supported by the U.S. Department of Energy, Office of Science, Basic Energy Sciences under Award \#DE-SC0016441.






\bibliographystyle{elsarticle-num}
\bibliography{MethodsX_Paper}

\begin{thebibliography}{10}
\expandafter\ifx\csname url\endcsname\relax
  \def\url#1{\texttt{#1}}\fi
\expandafter\ifx\csname urlprefix\endcsname\relax\def\urlprefix{URL }\fi
\expandafter\ifx\csname href\endcsname\relax
  \def\href#1#2{#2} \def\path#1{#1}\fi

\bibitem{Rosenbrock2017}
C.~W. Rosenbrock, E.~R. Homer, G.~Cs{\'{a}}nyi, G.~L.~W. Hart, {Discovering the
  building blocks of atomic systems using machine learning: application to
  grain boundaries}, npj Computational Materials 3 (2017) 29.
\newblock \href {http://dx.doi.org/10.1038/s41524-017-0027-x}
  {\path{doi:10.1038/s41524-017-0027-x}}.

\bibitem{Priedeman:2018structure}
J.~L. Priedeman, C.~W. Rosenbrock, O.~K. Johnson, E.~R. Homer, {Quantifying and
  Connecting Atomic and Crystallographic Grain Boundary Structure using Local
  Environment Representation and Dimensionality Reduction Techniques}, Acta
  Materialia (2018) Under Review.

\bibitem{bartok2013representing}
A.~P. Bart{\'o}k, R.~Kondor, G.~Cs{\'a}nyi, On representing chemical
  environments, Physical Review B 87~(18) (2013) 184115.

\bibitem{szlachta2014accuracy}
W.~J. Szlachta, A.~P. Bart{\'o}k, G.~Cs{\'a}nyi, Accuracy and transferability
  of gaussian approximation potential models for tungsten, Physical Review B
  90~(10) (2014) 104108.

\bibitem{de2015comparing}
S.~De, A.~P. Bart{\'o}k, G.~Cs{\'a}nyi, M.~Ceriotti, Comparing molecules and
  solids across structural and alchemical space, arXiv preprint
  arXiv:1601.04077.

\bibitem{PhysRevLett.104.136403}
A.~P. Bart\'ok, M.~C. Payne, R.~Kondor, G.~Cs\'anyi,
  \href{http://link.aps.org/doi/10.1103/PhysRevLett.104.136403}{Gaussian
  approximation potentials: The accuracy of quantum mechanics, without the
  electrons}, Phys. Rev. Lett. 104 (2010) 136403.
\newblock \href {http://dx.doi.org/10.1103/PhysRevLett.104.136403}
  {\path{doi:10.1103/PhysRevLett.104.136403}}.
\newline\urlprefix\url{http://link.aps.org/doi/10.1103/PhysRevLett.104.136403}

\bibitem{ISI:000175131400009}
S.~R. Bahn, K.~W. Jacobsen, An object-oriented scripting interface to a legacy
  electronic structure code, Comput. Sci. Eng. 4~(3) (2002) 56--66.
\newblock \href {http://dx.doi.org/10.1109/5992.998641}
  {\path{doi:10.1109/5992.998641}}.

\bibitem{Olmsted2009}
D.~L. Olmsted, S.~M. Foiles, E.~A. Holm, {Survey of computed grain boundary
  properties in face-centered cubic metals: I. Grain boundary energy}, Acta
  Materialia 57~(13) (2009) 3694--3703.
\newblock \href {http://dx.doi.org/10.1016/j.actamat.2009.04.007}
  {\path{doi:10.1016/j.actamat.2009.04.007}}.

\bibitem{Erickson}
H.~Erickson, E.~R. Homer, {Unpublished}.

\bibitem{Deringer2017}
V.~L. Deringer, G.~Cs{\'{a}}nyi, {Machine learning based interatomic potential
  for amorphous carbon}, Physical Review B 95~(9) (2017) 1--15.
\newblock \href {http://arxiv.org/abs/1611.03277} {\path{arXiv:1611.03277}},
  \href {http://dx.doi.org/10.1103/PhysRevB.95.094203}
  {\path{doi:10.1103/PhysRevB.95.094203}}.

\bibitem{olmsted2009survey}
D.~L. Olmsted, S.~M. Foiles, E.~A. Holm, Survey of computed grain boundary
  properties in face-centered cubic metals: I. grain boundary energy, Acta
  Materialia 57~(13) (2009) 3694--3703.

\bibitem{scikit-learn}
F.~Pedregosa, G.~Varoquaux, A.~Gramfort, V.~Michel, B.~Thirion, O.~Grisel,
  M.~Blondel, P.~Prettenhofer, R.~Weiss, V.~Dubourg, J.~Vanderplas, A.~Passos,
  D.~Cournapeau, M.~Brucher, M.~Perrot, E.~Duchesnay, Scikit-learn: Machine
  learning in {P}ython, Journal of Machine Learning Research 12 (2011)
  2825--2830.

\bibitem{CC01a}
C.-C. Chang, C.-J. Lin, {LIBSVM}: A library for support vector machines, ACM
  Transactions on Intelligent Systems and Technology 2 (2011) 27:1--27:27,
  software available at \url{http://www.csie.ntu.edu.tw/~cjlin/libsvm}.

\bibitem{Homer:2014hr}
E.~R. Homer, E.~A. Holm, S.~M. Foiles, D.~L. Olmsted,
  \href{http://link.springer.com/article/10.1007%2Fs11837-013-0801-2}{{Trends
  in grain boundary mobility: Survey of motion mechanisms}}, JOM-Journal Of The
  Minerals Metals {\&} Materials Society 66~(1) (2014) 114--120.
\newblock \href {http://dx.doi.org/10.1007/s11837-013-0801-2}
  {\path{doi:10.1007/s11837-013-0801-2}}.
\newline\urlprefix\url{http://link.springer.com/article/10.1007%2Fs11837-013-0801-2}

\bibitem{Homer:2013ce}
E.~R. Homer, S.~M. Foiles, E.~A. Holm, D.~L. Olmsted,
  \href{http://dx.doi.org/10.1016/j.actamat.2012.10.005}{{Phenomenology of
  shear-coupled grain boundary motion in symmetric tilt and general grain
  boundaries}}, Acta Materialia 61~(4) (2013) 1048--1060.
\newblock \href {http://dx.doi.org/10.1016/j.actamat.2012.10.005}
  {\path{doi:10.1016/j.actamat.2012.10.005}}.
\newline\urlprefix\url{http://dx.doi.org/10.1016/j.actamat.2012.10.005}

\bibitem{Chen:2016:XST:2939672.2939785}
T.~Chen, C.~Guestrin,
  \href{http://doi.acm.org/10.1145/2939672.2939785}{Xgboost: A scalable tree
  boosting system}, in: Proceedings of the 22Nd ACM SIGKDD International
  Conference on Knowledge Discovery and Data Mining, KDD '16, ACM, New York,
  NY, USA, 2016, pp. 785--794.
\newblock \href {http://dx.doi.org/10.1145/2939672.2939785}
  {\path{doi:10.1145/2939672.2939785}}.
\newline\urlprefix\url{http://doi.acm.org/10.1145/2939672.2939785}

\bibitem{lemaitre2016imbalanced}
G.~Lema\^{i}tre, F.~Nogueira, C.~K. Aridas,
  \href{http://arxiv.org/abs/1609.06570}{Imbalanced-learn: A python toolbox to
  tackle the curse of imbalanced datasets in machine learning}, CoRR
  abs/1609.06570.
\newline\urlprefix\url{http://arxiv.org/abs/1609.06570}

\bibitem{Han:2005:BNO:2141202.2141297}
H.~Han, W.-Y. Wang, B.-H. Mao,
  \href{http://dx.doi.org/10.1007/11538059_91}{Borderline-smote: A new
  over-sampling method in imbalanced data sets learning}, in: Proceedings of
  the 2005 International Conference on Advances in Intelligent Computing -
  Volume Part I, ICIC'05, Springer-Verlag, Berlin, Heidelberg, 2005, pp.
  878--887.
\newblock \href {http://dx.doi.org/10.1007/11538059_91}
  {\path{doi:10.1007/11538059_91}}.
\newline\urlprefix\url{http://dx.doi.org/10.1007/11538059_91}

\bibitem{Nguyen:2011:BOI:1972030.1972031}
H.~M. Nguyen, E.~W. Cooper, K.~Kamei,
  \href{http://dx.doi.org/10.1504/IJKESDP.2011.039875}{Borderline
  over\&\#45;sampling for imbalanced data classification}, Int. J. Knowl. Eng.
  Soft Data Paradigm. 3~(1) (2011) 4--21.
\newblock \href {http://dx.doi.org/10.1504/IJKESDP.2011.039875}
  {\path{doi:10.1504/IJKESDP.2011.039875}}.
\newline\urlprefix\url{http://dx.doi.org/10.1504/IJKESDP.2011.039875}

\end{thebibliography}






\newpage
\end{document}